\documentclass[11pt,a4paper]{article}
\usepackage{jcappub}
\usepackage[latin9]{inputenc}
\usepackage{textcomp}
\usepackage{amsmath}
\usepackage{amssymb}
\usepackage{color}

\makeatletter

\newcommand{\be}[1]{\begin{equation}\label{#1}}
\newcommand{\ee}{\end{equation}}
\newcommand{\ba}[1]{\begin{eqnarray}\label{#1}}
\newcommand{\ea}{\end{eqnarray}}
\newcommand{\rf}[1]{(\ref{#1})}
\newcommand{\nn}{\nonumber}

\makeatother

\begin{document}

\title{Coupled scalar fields in the late Universe: The mechanical approach and the late cosmic acceleration}

\author{Alvina~Burgazli$^{1}$,}
\author{Alexander~Zhuk$^1$,}
\author{Jo\~ao~Morais$^2$,}
\author{Mariam~Bouhmadi-L\'opez$^{3,4,2,5}$,}
\author{K.~Sravan~Kumar$^{3,4}$}

\emailAdd{aburgazli@gmail.com}
\emailAdd{ai.zhuk2@gmail.com}
\emailAdd{jviegas001@ikasle.ehu.eus}
\emailAdd{mbl@ubit.pt (On leave of absence from UPV/EHU and IKERBASQUE)}
\emailAdd{sravan@ubi.pt}

\affiliation{$^{1}$ Astronomical Observatory, Odessa National University, Dvoryanskaya st. 2, Odessa 65082, Ukraine\\}
\affiliation{$^2$ Department of Theoretical Physics, University of the Basque Country (UPV/EHU), P.O. Box 644, 48080 Bilbao, Spain\\}
\affiliation{$^{3}$Departamento de F\'{i}sica, Universidade da Beira Interior, Rua Marqu\^es D'\'Avila e Bolama, 6201-001 Covilh\~a, Portugal\\}
\affiliation{$^{4}$Centro de Matem\'atica e Aplica\c{c}\~oes da Universidade da Beira Interior (CMA-UBI), Rua Marqu\^es D'\'Avila e Bolama, 6201-001 Covilh\~{a}, Portugal\\}
\affiliation{$^{5}$IKERBASQUE, Basque Foundation for Science, 48011 Bilbao, Spain\\}

\date{Received: date / Accepted: date}

\abstract{%
In this paper, we consider the Universe at the late stage of its evolution and deep inside the cell of uniformity. At these scales, we consider the Universe to be filled with dust-like matter in the form of discretely distributed galaxies, a minimally coupled scalar field and radiation as matter sources. We investigate such a Universe in the mechanical approach. This means that the peculiar velocities of the inhomogeneities (in the form of galaxies) as well as fluctuations of other perfect fluids are non-relativistic. Such fluids are designated as coupled because they are concentrated around inhomogeneities. In the present paper we investigate the conditions under which a scalar field can become coupled, and show that, at the background level, such coupled scalar field behaves as a two component perfect fluid: a network of frustrated cosmic strings with EoS parameter $w=-1/3$ and a cosmological constant. The potential of this scalar field is very flat at the present time. Hence, the coupled scalar field can provide the late cosmic acceleration. The fluctuations of the energy density and pressure of this field are concentrated around the galaxies screening their gravitational potentials. Therefore, such scalar fields can be regarded as coupled to the inhomogeneities.%
}
\maketitle


\flushbottom

\section{\label{sec:1}Introduction}

\setcounter{equation}{0}

Under the assumption of homogeneity and isotropy for our Universe on its largest scales, the current observational data lead to the conclusion of the expansion of the Universe is accelerated,  as first implied by the type Ia supernova data almost twenty years ago \cite{SN1,SN2}. However, its nature is still a great challenge for modern cosmology. This phenomena has received the name dark energy (DE), which partly reflects its unclear nature.
The $\Lambda$CDM model, where the cosmological constant $\Lambda$ is responsible for the acceleration, is in very good agreement with observations \cite{7WMAP,9WMAP,Planck2013}. This model is equivalent to one with a perfect fluid with the constant EoS parameter $w=-1$. Unfortunately, this model has some puzzling and unresolved aspects such as the origin of $\Lambda$ and the coincidence problem \cite{Dolgov}. Therefore, a number of alternatives have been proposed which try to solve these problems, with models using scalar fields to explain DE being amongst the most popular ones\footnote{Besides being considered as DE candidates, scalar fields can also play the role of dark matter which could lead to the formation of gravitational structures in the earlier Universe. \cite{Matos1,Matos2}.}.
These are called quintessence \cite{quintess1,quintess2,quintess3} whenever $-1< w <0$, phantom \cite{phantom1,phantom2} when $ w <-1$, and quintom \cite{quintom} if there is $w=-1$ crossing.
One important and interesting feature of such models is that they can have a dynamical equation of state parameter which may solve the coincidence problem. However, there is also the possibility to construct models with constant $w$. This imposes severe restrictions on the form of the scalar field potential \cite{zhuk1996,ZBG}.

As we mentioned above, there is a number of alternative models of DE. Therefore, it is of great importance to propose a mechanism which can verify their viability. The theory of perturbations is a powerful tool to investigate cosmological models \cite{Mukhanov-book,Rubakov-book}. Such perturbations can be considered at any stage of the Universe evolution. In our paper, we investigate our Universe at the late stage of its evolution and deep inside of the cell of uniformity. At such scales the Universe looks highly inhomogeneous: galaxies, group and clusters of galaxies are already formed and can be considered as discrete sources for the gravitational potential. In our previous papers \cite{EZcosm2,EZcosm1}, we have shown that in this case the mechanical approach is an adequate tool to study scalar perturbations. In its turn, it enables us to get the gravitational potential and to consider the motions of galaxies \cite{EKZ2}. It is worth noting that similar ideas concerning the discrete cosmology have been discussed in the recent papers \cite{Ruth,EllGibb}. The mechanical approach was applied to a number of DE models to study their compatibility with the theory of scalar perturbations. For example, we considered the perfect fluid with a constant equation of state parameter \cite{BUZ1}, the model with quark-gluon nuggets \cite{Laslo2}, the CPL model \cite{CPL}, and the Chaplygin gas model \cite{Chaplygin}.

The mechanical approach works perfectly for the $\Lambda$CDM model where the peculiar velocities of the inhomogeneities (e.g. galaxies) can be considered as negligibly small (as compared with the speed of light), and, additionally, we consider scales deep inside of the cell of uniformity\footnote{\label{foot1}For an estimation of the dimension of the cell of uniformity (the scale of homogeneity) from the point of view of the gravitational interaction features at large distances, see \cite{Ein1}.}. Then, we may drop the peculiar velocities at the first order approximation \cite{EZcosm2,EZcosm1}. As we mentioned above, such an approach was generalised also to the case of cosmological models with different perfect fluids which can play the role of dark energy and dark matter.
Fluctuations of these additional perfect fluids also form their own inhomogeneities. In the mechanical approach, it is supposed that the velocities of displacement of such
inhomogeneities is of the order of the peculiar velocities of inhomogeneities of dust-like matter, i.e. they are
non-relativistic. In some sense, these two types of inhomogeneities are ``coupled'' to each other \cite{coupled}. This is an important point. This means that for the considered models, we investigate the possibility of the existence of such ``coupled" fluids\footnote{\label{foot2}In what follows, we shall omit quotation marks for the word ``coupled".}. They can play an important role of dark matter and can be distributed around the baryonic inhomogeneities (e.g., galaxies) in such a way that it can solve the problem of flatness of the rotation curves \cite{Laslo2}. It is worth noting that the generalization of the mechanical approach to the case of non-zero peculiar velocities and for all cosmological scales was performed in the recent papers \cite{Ein1,Ein2}. This generalization includes also the case of uncoupled perfect fluids.

In the present paper, we consider a cosmological model with a scalar field minimally coupled to gravity. The Universe is also filled with dust-like matter (in the form of discrete galaxies and group of galaxies) and radiation. We study the theory of scalar perturbations for such model and obtain a condition under which the inhomogeneities of dust-like matter and the inhomogeneities of the scalar field can be coupled to each other (in the sense pointed above). We demonstrate that this condition imposes rather strong restrictions on the scalar field itself. First, the coupled scalar field behaves (at the background level) as a two-component perfect fluid: a cosmological constant and a network of frustrated cosmic strings. The latter has a parameter of EoS $w=-1/3$. The potential of such scalar field is very flat at the present time (see Fig. 1). This flatness condition is a natural consequence of the current acceleration of the Universe, as the contribution of the term with $w=-1/3$ has to be very small at present, as implied by the current observations. Second, the fluctuations of the scalar field are absent and the energy density and pressure of the scalar field fluctuate due to the interaction of the gravitational potential with the scalar field background. Nevertheless, such a coupled scalar field is in concordance with the theory of scalar perturbations (at relatively small cosmological scales \cite{Ein1}) and contributes to the gravitational potential. The fluctuations of the energy density of the scalar field are concentrated around the galaxies, screening their gravitational potentials. Such a distribution of the energy density of the scalar field fluctuations justifies the coupling condition. We obtain the expressions for the gravitational potential for flat, open and closed topologies of the Universe. An important property of this potential is that its averaged (over the volume of the Universe) value is equal to zero, as it should be \cite{BUZ1,Ein1}. Consequently, the averaged value of the energy density fluctuations of scalar field is also equal to zero. We also determine the form of the scalar field potential.

The paper is structured as follows. The background equations are given in Sec.~\ref{sec:2}. In Sec.~\ref{sec:3}, we investigate within the mechanical approach the scalar perturbations for the considered cosmological model. We define the conditions under which the scalar field satisfies the equations obtained. Here, we also obtain the equation for the non-relativistic gravitational potential and its solutions. In Sec.~\ref{sec:4}, we determine the form of the scalar field potential. The main results are summarised in the concluding Sec.~\ref{sec:5}.

\section{\label{sec:2}Background equations}

\setcounter{equation}{0}

To start with, we consider the background equations which describe the homogeneous and isotropic Universe. The background metric is the Friedmann-Lema\^itre-Ro-bertson-Walker one
\be{2.1}
	ds^2=a^2(\eta)\left(d\eta^2-\gamma_{\alpha\beta}dx^{\alpha}dx^{\beta}\right)
	\,,
\ee
where the exact form of the metric $\gamma_{\alpha\beta}$ is defined by the topology of the Universe. For generality, we shall consider all three possible topologies with scalar curvatures $\mathcal K=-1,0,+1$ for open, flat and closed Universes, respectively.
The Universe is filled with a scalar field minimally coupled to gravity. Such a field is described by the action
\be{2.2}
	S_\phi=\int d^4 x\sqrt{-g}\left[\frac{1}{2}g^{\mu\nu}\partial_\mu\phi\,\partial_\nu\phi-V(\phi)\right]
\ee
and has the energy-momentum tensor:
\be{2.3}
	T_\nu^\mu(\phi) = g^{\mu\lambda}\partial_\nu\phi\partial_\lambda\phi
	-\delta^\mu_\nu\left[\frac{1}{2}g^{\lambda\rho}\partial_\lambda\phi\,\partial_\rho\phi-V(\phi)\right]
	\, .
\ee
The equation of motion reads
\be{2.4}
	\frac{1}{\sqrt{-g}}\partial_\mu\left(\sqrt{-g}g^{\mu\nu}\partial_\nu\phi\right)+\frac{dV}{d\phi}(\phi)=0
	\, .
\ee
For the homogeneous and isotropic Universe, the scalar field depends only on time. Let $\phi_c (\eta)$ describes such a background scalar field. Then, for the background energy density and pressure we get
\ba{2.5}
	\bar{T}^0_0\equiv \bar \varepsilon_{\varphi} &=& \frac{1}{2a^2}(\phi'_c)^2+V(\phi_c)
	\, ,
	\\
	\label{2.6}
	-\bar{T}^i_i\equiv \bar p_{\varphi} &=& \frac{1}{2a^2}(\phi'_c)^2-V(\phi_c)
	\, ,
\ea
where the prime denotes the derivative with respect to the conformal time $\eta$. The background equation of motion is
\be{2.7}
	\phi''_c+2\mathcal{H}\phi'_c+a^2\frac{dV}{d\phi}(\phi_c)=0
	\, ,
\ee
where $\mathcal{H}=a'/a$.

As matter sources, we also include dust-like matter (baryonic and CDM) and radiation.  The background (i.e. average) energy density of the dust-like matter takes the form $\bar\varepsilon_{\mathrm{dust}} =\bar \rho c^2/a^3$, where $\bar \rho=\mbox{const}$ is the average comoving rest mass density \cite{EZcosm1}. As usual, for radiation, we have the EoS $\bar p_{\mathrm{rad}}=(1/3)\bar \varepsilon_{\mathrm{rad}}$ and $\varepsilon_{\mathrm{rad}} \sim 1/a^4$.

For the above described cosmological model, the Friedmann and Raychaudhuri equations take, respectively, the form
\ba{2.8}
	\mathcal{H}^2 = \frac{\kappa a^2}{3}\left[\bar\varepsilon_\mathrm{dust}+\bar\varepsilon_\mathrm{rad}+\frac{1}{2}(\phi'_c)^2/a^2+V(\phi_c)\right]-\mathcal{K}
\ea
and
\be{2.9}
	\mathcal{H}'=\frac{1}{3}a^2\kappa\left[-\bar\varepsilon_\mathrm{rad}-\frac{1}{2}\bar\varepsilon_\mathrm{dust}-(\phi'_c)^2/a^2+V(\phi_c)\right]
	\,,
\ee
where $\kappa\equiv 8\pi G_N/c^4$ ($c$ is the speed of light and $G_N$ is the Newtonian gravitational constant) and we have used Eqs.~\rf{2.5} and \rf{2.6}.


\section{\label{sec:3}Scalar perturbations}

\setcounter{equation}{0}

Let us turn now to the scalar perturbations. Then, the metrics reads \cite{Mukhanov-book,Rubakov-book}:
\be{3.1}
	ds^2=a^2(\eta)\left[\left(1+2\Phi\right)d\eta^2-\left(1-2\Psi\right)\gamma_{\alpha\beta}dx^{\alpha}dx^{\beta}\right]
	\, .
\ee
The perturbations of the scalar field energy-momentum tensor are \cite{Rubakov-book}:
\ba{3.2}
	\delta T^0_0 &\equiv& \delta\varepsilon_\varphi=-\frac{1}{a^2}\left(\phi'_c\right)^2\Phi+\frac{1}{a^2}\phi'_c\varphi' + \frac{dV}{d\phi}(\phi_c)\varphi
	\, ,
	\\
	\delta T^0_i &=& \frac{1}{a^2}\phi'_c\partial_i\varphi\label{3.3}
	\, ,
	\\
	\delta T_j^i&\equiv&-\delta^i_j\delta p_\varphi
	\, ,
	\nn\\
	\delta p_\varphi&=&-\frac{1}{a^2}\left(\phi'_c\right)^2\Phi+\frac{1}{a^2}\phi'_c\varphi'-\frac{dV}{d\phi}(\phi_c)\varphi\label{3.4}
	\, ,
\ea
where we split the scalar field into its background part $\phi_c (\eta)$ and its fluctuations part $\varphi(\eta,\vec{r})$:
\be{3.5}
	\phi=\phi_c+\varphi
	\, .
\ee

For the considered model, the Einstein equations are reduced (after linearising the system of 3 equations) to:
\ba{3.6}
	\Delta\Phi - 3\mathcal{H}(\Phi'+\mathcal{H}\Phi)+3\mathcal{K}\Phi
	&=&\frac{\kappa}{2}a^2\left(\delta\varepsilon_{\mathrm{dust}}+ \delta\varepsilon_{\mathrm{rad}}\right) 
	\nn\\
	&{}&~
	- \frac{\kappa}{2}\left[(\phi'_c)^2\Phi-\phi'_c\varphi'-a^2\frac{dV}{d\phi}(\phi_c)\varphi\right]
	\, ,
\ea
\be{3.7}
	\partial_i \Phi'+\mathcal{H}\partial_i \Phi=\frac{\kappa}{2}\phi'_c\partial_i\varphi
\ee
and
\ba{3.8}
	&&\frac{2}{a^2}\left[\Phi''+3\mathcal{H}\Phi'+\Phi\left(2\frac{a''}{a}-\mathcal{H}^2-\mathcal{K}\right)\right]
	\nn\\
	&&
	\hspace{4cm} 
	=\kappa\left[ \delta p_{\mathrm{rad}} - \frac{1}{a^2}\left(\phi'_c\right)^2\Phi+\frac{1}{a^2}\phi'_c\varphi'-\frac{dV}{d\phi}(\phi_c)\varphi\right]
	\,.
\ea
Here, according to the mechanical approach (see details in \cite{EZcosm2,EZcosm1}), we drop the terms containing the peculiar velocities of the inhomogeneities and radiation as these are negligible when compared with their respective energy density and pressure fluctuations. However, such comparison with respect to the scalar field is not evident since the quantity treated as the peculiar velocity of the scalar field is proportional to the scalar field perturbation $\varphi$. Therefore, in our analysis we propose the following strategy. First, we preserve the scalar field perturbation in \rf{3.7} since we keep it in Eqs. \rf{3.6} and \rf{3.8}. Then, a subsequent analysis of the equations must show whether or not we can equate to zero the right hand side (r.h.s.) of Eq. \rf{3.7}. In what follows, we shall demonstrate that for the coupled scalar field the r.h.s. of this equation can indeed be set to zero in a consistent way within the mechanical approach as it usually happens for the coupled fluids \cite{coupled}.
We also applied the standard reasoning to put $\Psi=-\Phi$ \cite{Rubakov-book}, i.e., we have assumed absence of anisotropies. In Eq.~\rf{3.6}, $\Delta$ is the Laplace operator with respect to the metric $\gamma_{\alpha\beta}$. From Eq.~\rf{3.7}, we get the following relation:
\be{3.9}
	\Phi'+\mathcal{H} \Phi=\frac{\kappa}{2}\phi'_c\varphi
	\, .
\ee

Let us consider now Eq.~\rf{3.8} in more detail. The substitution of $\Phi'$ from \rf{3.9} into \rf{3.8} gives
\ba{3.10}
	\Phi\left[ \mathcal{H}'-\mathcal{H}^2-\mathcal{K}+\kappa\frac{1}{2}(\phi'_c)^2 \right] 
	= \varphi\left[-\frac{\kappa}{2}\phi''_c-\mathcal{H}\kappa\phi'_c-\frac{a^2}{2}\kappa\frac{dV}{d\phi}(\phi_c) \right]+\kappa\frac{a^2}{2}\delta p_{\mathrm{rad}}
	\, ,
\ea
where we have also used the relation $2a''/a=2(\mathcal{H}' + \mathcal{H}^2)$. With the help of the Eqs.~\rf{2.8} and \rf{2.9}, this equation finally takes the form
\be{3.11}
	\Phi\left[ -\frac{2}{3}a^2\kappa \bar\varepsilon_\mathrm{rad}-\frac{1}{2}a^2\kappa\bar\varepsilon_\mathrm{dust} \right]=\kappa\frac{a^2}{2}\delta p_{\mathrm{rad}}
	\, ,
\ee
where we have taken into account the equation of motion \rf{2.7}. Because, $\bar\varepsilon_\mathrm{rad}\sim 1/a^4$ and $\bar\varepsilon_\mathrm{dust}\sim 1/a^3$, we can drop the first term in the brackets in the left-hand-side of this equation and obtain
\be{3.12}
\delta p_\mathrm{rad}=-\Phi\bar\varepsilon_\mathrm{dust}=-\Phi\frac{\bar\rho c^2}{a^3} = \frac13 \delta \varepsilon_\mathrm{rad}
\,,
\ee
similar to the expression (4.19) in \cite{EZcosm2}. Therefore, the spatial distribution of $\delta \varepsilon_\mathrm{rad}$ is defined by the gravitational potential $\Phi$ derived below in Eq. \rf{3.32}.

Now, we turn to Eq.~\rf{3.6}. Since (see \cite{EZcosm1})
\be{3.13}
	\delta\varepsilon_{\mathrm{dust}}=\frac{\delta\rho c^2}{a^3}+\frac{3\bar\rho\, \Phi}{a^3}
	\, ,
\ee
where $\delta\rho$ is the difference between real and average rest mass densities for the dust-like matter:
\be{3.14}
	\delta\rho=\rho -\bar\rho
\ee
and taking into account \rf{3.12}, this equation reads
\ba{3.15}
	\Delta\Phi-3\mathcal{H}(\Phi'+\mathcal{H}\Phi)+3\mathcal{K}\Phi
	=\frac{\kappa}{2} \frac{\delta\rho c^2}{a} 
	- \frac{\kappa}{2}\left[(\phi'_c)^2\Phi-\phi'_c\varphi'-a^2\frac{dV}{d\phi}(\phi_c)\varphi\right]
	\,.
\ea
From \rf{3.9} we get
\ba{3.16}
	\varphi&=&\frac{\Phi'+\mathcal{H}\Phi}{\frac{\kappa}{2}\phi'_c}
	\, ,
	\\
	\varphi'&=&\frac{\Phi''+\mathcal{H}'\Phi+\mathcal{H}\Phi'}{\frac{\kappa}{2}\phi'_c}
	-\frac{\Phi'+\mathcal{H}\Phi}{\frac{\kappa}{2}(\phi'_c)^2}\phi''_c
	\, .
	\label{3.17}
\ea
The substitution of \rf{3.16} and \rf{3.17} into \rf{3.15} gives:
\ba{3.18}
	\Delta\Phi-\frac{\kappa}{2}\frac{\delta\rho c^2}{a} &=&\Phi\left[3\mathcal{H}^2-3\mathcal{K}-\frac{\kappa}{2}(\phi'_c)^2+\mathcal{H}'-\mathcal{H}\frac{\phi''_c}{\phi'_c}+a^2\frac{dV}{d\phi}(\phi_c)\frac{1}{\phi'_c}\mathcal{H}\right]
	\nn\\
	&&~+\Phi'\left[4\mathcal{H}-\frac{\phi''_c}{\phi'_c}+a^2\frac{dV}{d\phi}(\phi_c)\frac{1}{\phi'_c}\right] +\Phi''
	\, .
\ea
Since
\be{3.19}
	\Phi'=\frac{d\Phi}{da}a\mathcal{H}
	\,,
	\qquad
	\Phi'' = \frac{d^2\Phi}{da^2}a^2\mathcal{H}^2+\frac{d\Phi}{da}a\mathcal{H}'+\frac{d\Phi}{da}a\mathcal{H}^2
	\, ,
\ee
then Eq.~\rf{3.18} can be written in the form
\ba{3.20}
	\Delta\Phi-\frac{\kappa}{2}\frac{\delta\rho c^2}{a}
	&=& \Phi\left[3\mathcal{H}^2-3\mathcal{K}-\frac{\kappa}{2}(\phi'_c)^2+\mathcal{H}'-\mathcal{H}\frac{\phi''_c}{\phi'_c}+a^2\frac{dV}{d\phi}(\phi_c)\frac{1}{\phi'_c}\mathcal{H}\right]
	\nn \\
	&&~+\frac{d\Phi}{da}a\left[5\mathcal{H}^2+\mathcal{H}'
	-\mathcal{H}\frac{\phi''_c}{\phi'_c}+a^2\frac{dV}{d\phi}(\phi_c)\frac{1}{\phi'_c}\mathcal{H}\right]
	+\frac{d^2\Phi}{da^2}\mathcal{H}^2a^2
	\, ,
\ea
which after substitution $\Phi =\Omega/a$, where $\Omega$ is a function of $a$ and the spatial coordinates, reads
\ba{3.21}
	\frac{\Delta\Omega}{a}-\frac{\kappa}{2}\frac{\delta\rho c^2}{a}&=& -\frac{\Omega}{a}\left[3\mathcal{K}+\frac{\kappa}{2}(\phi'_c)^2\right]
	\nn\\
	&&~+\frac{d\Omega}{da}\left[3\mathcal{H}^2+\mathcal{H}'-\mathcal{H}\frac{\phi''_c}{\phi'_c}+a^2\frac{dV}{d\phi}(\phi_c)\frac{1}{\phi'_c}\mathcal{H}\right]
	+\frac{d^2\Omega}{da^2}a\mathcal{H}^2
	\, .
\ea
We can use this equation to determine the unknown function $\Omega$ and, consequently, the gravitational potential $\Phi$.

In what follows, the dust like matter component is considered in the form of discrete distributed inhomogeneities (e.g. galaxies and their group and clusters). Then, we are looking for solutions of \rf{3.21} which have a Newtonian limit near gravitating masses. Such an asymptotic behaviour  will take place if we impose{\footnote{\label{foot3}It can be easily seen that the condition $\Omega = \Omega (\vec{r})$ demands also $(\phi'_c)^2=\mathrm{const}$ and vice versa}} $\Omega = \Omega (\vec{r})$ (see solutions \rf{3.32} and \rf{3.32a} below). Moreover, such a choice is in agreement with the transition to the astrophysical approach (in other words, the Minkowski background limit) where $a \rightarrow \mathrm{const} \Rightarrow \mathcal{H} \rightarrow 0$ and all background energy densities are equal to zero (e.g. this means that $\bar \rho =0$ and $\phi'_c =0$), and we should select the flat topology $\mathcal{K}=0$. In this limit, if the dust-like matter is described by the discrete distributed gravitating sources (e.g. galaxies) with masses $m_i$ and the rest-mass density
\be{3.22}
	\rho = \sum_i m_i\delta (\vec{r}-\vec{r}_i)
	\, ,
\ee
the gravitational potential $\Phi$ is
\be{3.23}
	\Phi = - \frac{G_N}{c^2 }\frac{1}{a}\sum_i\frac{m_i}{|\vec{r}-\vec{r}_i|}= - \frac{G_N}{c^2 }\sum_i\frac{m_i}{|\vec{R}-\vec{R}_i|}
	\, ,
\ee
as it should be \cite{Landau}. In Eq.~\rf{3.23}, we took into account the relations between the physical and comoving radius vectors:
$\vec{R} = a\vec{r}$. This equation also demonstrates that $\Phi \sim 1/a$.

Now, we analyse the case $\Omega = \Omega (\vec{r}) \Rightarrow \Phi \sim 1/a$ and $(\phi'_c)^2=\mathrm{const}$ in more details.
Let us denote $\phi'_c=\beta=\mathrm{const}$. Then we get
\be{3.24}
	\phi_c=\beta\eta+\gamma,\quad \gamma=\mathrm{const}
	\, .
\ee
The substitution of \rf{3.24} into the equation of motion \rf{2.7} gives:
\ba{3.25}
	2\frac{a'}{a}\beta&+&a^2\frac{d V}{d\phi}(\phi_c)=2\frac{a'}{a}\beta+a^2\frac{V'}{\beta}=0\, , \quad\Rightarrow\quad V=\frac{\beta^2}{a^2}+ V_\infty,\quad V_\infty=\mathrm{const}
	\, .
\ea
Obviously, $V_\infty$ plays the role of the cosmological constant. It appears as a solution of the equations of motion. To achieve the accelerated expansion of the late Universe, there is no need to include the cosmological constant in the model by hand. This is an important consequence of our approach. To get the late cosmic acceleration, we must demand that $V_\infty > 0$. Using the data of the Planck mission in combination with other experiments \cite{Planck2013}, we obtain that the potential $V$ is very flat at the present time (see Fig. 1 below). Therefore, the coupled scalar field can provide the late cosmic acceleration.

For the background energy density and pressure (see Eqs.~\rf{2.5} and \rf{2.6}) we obtain:
\be{3.26}
	\bar \varepsilon_{\varphi} = \frac{3}{2}\frac{\beta^2}{a^2}+V_\infty
	\,,
	\qquad
	\bar p_{\varphi} = -\frac{1}{2}\frac{\beta^2}{a^2}-V_\infty
	\, .
\ee
It can be easily seen that the considered scalar field behaves (at the background level) as a two-component perfect fluid: a cosmological constant and a network of frustrated cosmic strings \cite{ZBG,BUZ1,Chaplygin-inflation}. The latter has the parameter of EoS $w=-1/3$.
This behaviour is a consequence of imposing the coupling between the scalar field perturbations and the inhomogeneities. In other words, we derived a specific form for the time dependence of the background scalar field and for its potential (see Eqs. \rf{3.24} and \rf{3.25}, respectively) so that the scalar field is consistent with such a coupling to the inhomogeneities. Obviously, in the general case, when the coupling condition is not imposed, the scalar field is not limited to these specific solutions.

There is also another very important feature of the considered scalar field. Since $\Phi \sim 1/a$, then we find $\Phi'+ \mathcal{H} \Phi =0$. Hence, it follows from Eqs.~\rf{3.16} and \rf{3.17} that fluctuations of the scalar field are absent: $\varphi = \varphi' =0$. The physical reason of this is that the ``coupling'' between the inhomogeneities of the dust-like matter and of the scalar field imposes a strong restriction on the scalar field. Nevertheless, the above analysis demonstrates that such a scalar field can exist, i.e. such scalar field does not contradict the above equations both at the background and at the linear perturbation levels. On the other hand, the fluctuations of the energy density and pressure of the scalar field are non-zero (see Eqs. \rf{3.2} and \rf{3.4}):
\be{3.27}
	\delta\varepsilon_\varphi=\delta p_\varphi=-\frac{1}{a^2}\left(\phi'_c\right)^2\Phi = -\frac{\beta^2}{a^3} \Omega (\vec{r}) \neq 0
	\, .
\ee
These fluctuations arise due to the interaction between the scalar field background and the gravitational potential. It is well known that the energy densities and pressure but not fields are measurable values.
Eq. \rf{3.27} shows that $\delta\varepsilon_{\varphi}\sim 1/a^3$ in analogy also with the fluctuations of the energy density for a perfect fluid with the constant equation of state parameter $\omega =-1/3$ \cite{BUZ1}. However, there is also a difference between the considered scalar field and such a perfect fluid. For the scalar field $\delta\varepsilon_{\varphi}=\delta p_{\varphi}$ and the EoS parameter for the fluctuations is $\delta p_{\varphi}/\delta\varepsilon_{\varphi}=1$ whereas for the fluctuations of the perfect fluid the EoS parameter is still -1/3 as for the background matter. Therefore, these two models are not completely equivalent to each other.

Additionally, the fluctuations of the energy density of the scalar field contribute to the gravitational potential. To prove it, we can rewrite Eq.~\rf{3.21} as follows (for the considered case $\Omega = \Omega (\vec{r})$):
\be{3.28}
	\Delta\Omega + 3\mathcal{K}\Omega =\frac{\kappa}{2}\left(\delta\rho c^2 +\widetilde{\delta\varepsilon_\varphi}\right)
	\, ,
\ee
where we have introduced the comoving fluctuations of the energy density of scalar field:
\be{3.29}
	\widetilde{\delta\varepsilon_\varphi}=a^3 \delta\varepsilon_\varphi = -\left(\phi'_c\right)^2\Omega =-\beta^2\Omega
	\, .
\ee
Eq.~\rf{3.28} can also be written in the form
\be{3.30}
	\Delta f - \lambda^2 f = 4\pi G_N\delta\rho
	\, ,
\ee
where $f(\vec{r})=c^2\Omega (\vec{r})$ and
\be{3.31}
	\lambda^2 = -\frac{\kappa}{2}\beta^2 - 3\mathcal{K}
	\, .
\ee
This equation can be solved for any spatial topology \cite{BUZ1}. For example, in the case of spatially flat ($\mathcal{K}=0$) and hyperbolic ($\mathcal{K}=-1$) geometries
we get, respectively:
\ba{3.32}
	f&=&-G_{N}\underset{i}{\sum}\frac{m_{i}}{\vert\mathbf{r}-\mathbf{r}_{i}\vert}\exp\left(-\lambda\vert\mathbf{r}-\mathbf{r}_{i}\vert\right)+\frac{4\pi G_{N}\bar{\rho}}{\lambda^{2}}
	\, ,
	\\
\label{3.32a}
	f&=&-G_N\sum_i m_i\frac{\exp(-l_i\sqrt{\lambda^2+1}\, )}{\sinh l_i}+\frac{4\pi G_N\overline\rho}{\lambda^2}\, ,\quad 0 < l_i <+\infty
	\, ,
\ea
where $l_i$ denotes the geodesic distance between the $i$-th mass $m_i$ and the point of observation. To obtain these physically reasonable  solutions (i.e. solutions which have the correct Newtonian limit near the inhomogeneities and converge at spatial infinity), we impose that\footnote{In the flat case this condition should be replaced by $\lambda>0$} $\lambda^2 >0$. In the case of spatially flat topology, this means that $\beta^2<0$. Eq. \rf{3.26} shows that the scalar field background energy density can be negative if $V_\infty<-\beta^2/a^2$. To avoid a possible problem with a ghost-like instability, we can impose the condition\footnote{This condition defines a minimum scale factor $a_\mathrm{min}=|\beta|/\sqrt{V_\infty}$ such that for $a>a_\mathrm{min}$ the energy density of the scalar field is positive.} $V_\infty>-\beta^2/a^2$.
Moreover, for a hyperbolic space, $\lambda^2$ can acquire positive values if $\beta^2$ is positive and such possible problem is absent for sure.

In the case of spherical spatial topology $(\mathcal{K}=+1)$ and for the physical reasonable values $\beta^2>0$, we get $\lambda^2<0$. Then, the solution of Eq. \rf{3.30} is \cite{BUZ1}:
\ba{3.32b}
	f=-G_N\sum_i m_i\frac{\sin\left[(\pi-\l_i)\sqrt{1-\lambda^2}\, \right]}{\sin\left(\pi\sqrt{1-\lambda^2}\, \right)\sin{l_i}}+\frac{4\pi G_N\overline\rho}{\lambda^2}\, , \quad 0<l_i\leq \pi
	\, .
\ea
For $\sqrt{1-\lambda^2} \neq 2,3,\ldots$, this formula is finite at any point $l_i \in (0,\pi]$ and has the Newtonian limit for $l_i\to 0$.

As it follows from \rf{3.27}, {\it the fluctuations of the density of the scalar field are concentrated around the inhomogeneities of the dust-like matter (i.e. around the galaxies) which is in full agreement with the coupling condition.} The presence of these fluctuations leads to the screening of the gravitational potential as it follows from Eqs. \rf{3.32}-\rf{3.32b}.
It was also shown in \cite{BUZ1} that for all topologies of the space the solutions found for the gravitational potential satisfy the important condition that the total gravitational potential averaged over the whole Universe is equal to zero: $\bar f =0 \Rightarrow \overline \Omega =0$. This demand results in another physically reasonable condition: $\overline{\delta\varepsilon_{\varphi}}=0$ (see Eq.~\rf{3.27}). It is obvious that the averaged value of the fluctuations should be equal of zero.


\section{\label{sec:4}Scalar field potential}

\setcounter{equation}{0}

In the previous section we have obtained the dependence of the scalar field potential compatible with the mechanical approach (see Eq.~\eqref{3.25}). We now seek to obtain the shape of $V(\phi_c)$ as a function of the scalar field itself. In order to do this, we begin by writing the Friedmann equation as
\ba{4.1}
	\mathcal{H}^2 &=& \frac{H_0^2}{c^2}a^2
	\left[
		\Omega_{V_\infty}
		+ \left(\Omega_\mathcal{K}+\Omega_\beta\right) \left(\frac{a_0}{a}\right)^2
		+\Omega_\mathrm{dust}\left(\frac{a_0}{a}\right)^3
		+\Omega_\mathrm{rad}\left(\frac{a_0}{a}\right)^4
	\right],
\ea
where
\ba{4.2}
	\Omega_{V_\infty} &\equiv& \frac{\kappa c^2 V_\infty}{3H_0^2}
	\,,
	\nn\\
	\Omega_\mathcal{K} &\equiv& -\frac{\mathcal{K}c^2}{H_0^2a_0^2}
	\,,
	\nn\\
	\Omega_{\beta} &\equiv& \frac{\kappa c^2 \beta^2}{2H_0^2a_0^2}\,,
	\nn\\
	\Omega_{\textrm{dust}} &\equiv& \frac{\kappa c^2 \bar\varepsilon_{\textrm{dust},0}}{3H_0^2}= \frac{\kappa c^4 \bar\rho_c}{3H_0^2a_0^3}\,,
	\nn\\
	\Omega_\textrm{rad} &\equiv& \frac{\kappa c^2 \bar\varepsilon_{\textrm{rad},0}}{3H_0^2}\,.
\ea
After some algebra and using Eq.~\eqref{3.24} to replace the conformal time $\eta$ by $\phi_c$, we can integrate Eq.~\eqref{4.1} from some initial value $a_i$ till $a_f$ and obtain
\ba{4.3}
	\phi_c(a_f)-\phi_c(a_i)=\pm A\int_{\frac{a_i}{a_0}}^{\frac{a_f}{a_0}}\frac{d (a/a_0)}{
	\sqrt{
	\left(\frac{a}{a_0}\right)^4
	+ c_2
	\left(\frac{a}{a_0}\right)^2
	+ c_1
	\frac{a}{a_0}
	+c_0
	}}
	\,.
\ea
Here the dimensionless coefficients $c_i$'s are defined as $c_2\equiv (\Omega_\mathcal{K}+\Omega_{\beta})/\Omega_{V_\infty}$, $c_1 \equiv \Omega_\textrm{dust}/\Omega_{V_\infty}$, $c_0 \equiv \Omega_\textrm{rad}/\Omega_{V_\infty}$, and the constant $A$ is
\ba{4.4}
	 A \equiv \sqrt{\frac{2}{\kappa}\frac{\Omega_\beta}{\Omega_{V_\infty}}}
	\,.
\ea

It follows from current observations \cite{Planck2013} that at the present time the main contributions to the total energy density come from the cosmological constant and dust, i.e., $|c_2|\ll c_1<1$. Furthermore, since $c_{0}>0$, $c_{1}>0$, and the integrand on the r. h. s. of the equation behaves as $1/(a/a_0)^2$ when $a/a_0\gg1$, we find that the integral is well defined and finite for all values of $a_{i,f}\in[0,+\infty)$, as well as in the limit of $a_f\rightarrow+\infty$. This means that from the distant past, $a=0$, till the distant future, $a\rightarrow+\infty$, the variation of the scalar field, $\Delta\phi_c=|\phi_c(+\infty)-\phi_c(0)|$, is finite.

Using the freedom in the definition of the integration constants, we can take the limit $a_f\rightarrow+\infty$, set $\phi_c(+\infty) = 0$, and rewrite Eq.~\eqref{4.3} as
\ba{4.5}
	 &&\phi_c(a)
	= \pm A\int_{\frac{a}{a_0}}^{+\infty}\frac{d ( a/a_0)}{
	\sqrt{
	\left(\frac{ a}{a_0}\right)^4
	+ c_2
	\left(\frac{ a}{a_0}\right)^2
	+ c_1
	\frac{ a}{a_0}
	+c_0
	}}
	\,.
\ea
The solution with a $+$ $(-)$ sign can then be identified with the case where the scalar field rolls down the potential to the left (right) of the limiting value $\phi_c(+\infty)$, i.e. with negative (positive) values of $\phi_c$.
The relation \eqref{4.5} can be inverted numerically and inserted in Eq.~\eqref{3.25} in order to obtain $V(\phi_c)$.
In the future, when $a/a_0\gg1$, we can take the approximation that only the quartic term inside the square root contributes to the integral in Eq.~\eqref{4.5}. In this case we find that the potential is given by
\ba{4.6}
	 V \simeq {V_\infty} + \frac{\beta^2}{a_0^2}\left(\frac{\phi_c}{A}\right)^2
	= V_\infty\left[
		1
		+\frac{2\Omega_\beta}{3\Omega_\lambda}\left(\frac{\phi_c}{A}\right)^2
	\right]
	\,.
\ea

\begin{figure}[t]
\centering
	\includegraphics[width=.7\columnwidth]{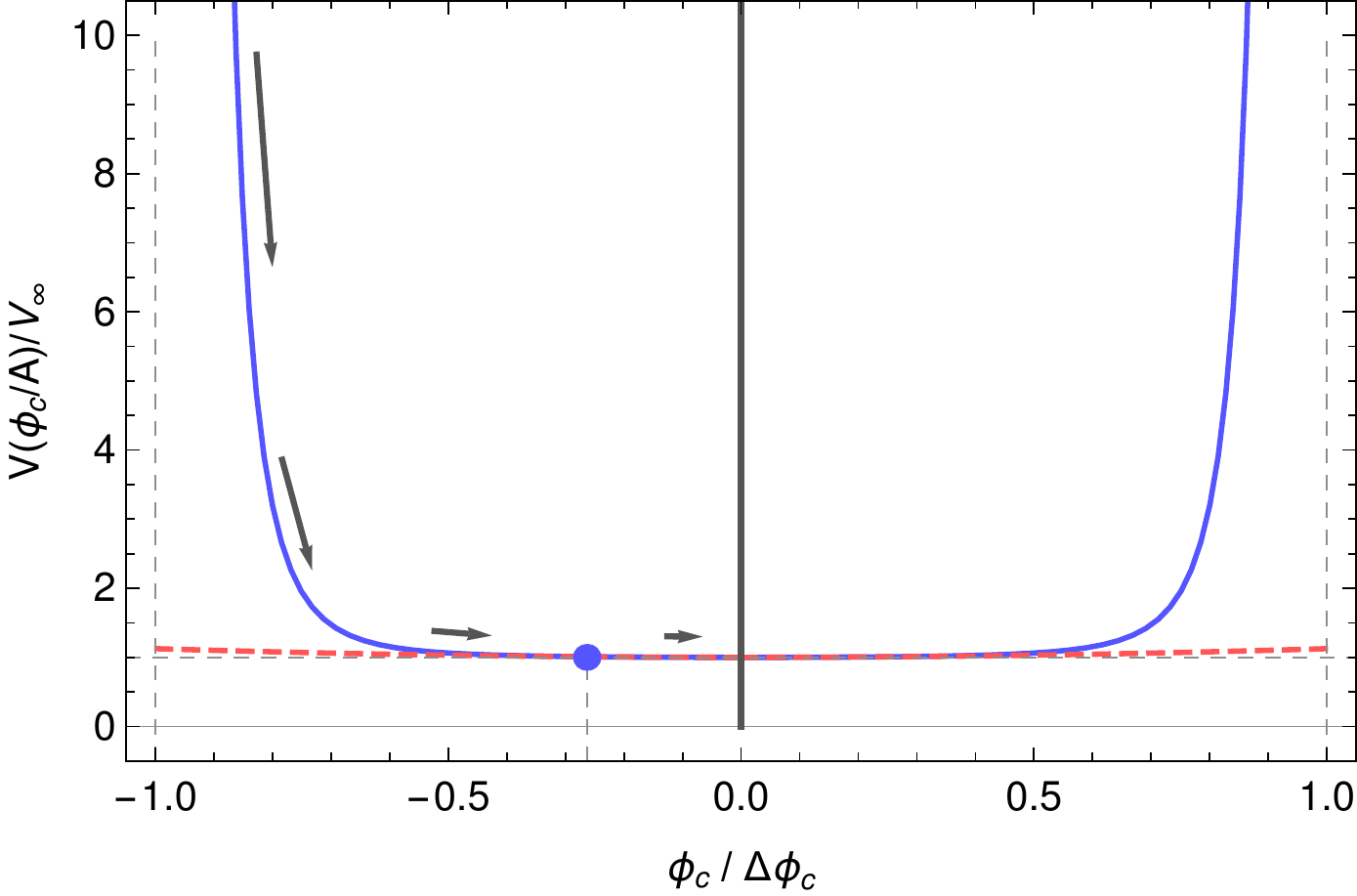}
	\caption{
		\label{plot1}
		The full scalar field potential obtained from Eq.~\eqref{4.5} (blue full curve) and the late time quadratic approximation (red dashed curve) as functions of $\phi_c/\Delta\phi_c$.
		If initially at a value $\phi_c=-\Delta\phi_c$, the scalar field rolls down the potential, as indicated by the black arrows, until it reaches the minimum of the potential at $\phi_c=0$. Since this point corresponds to the distant future $(a\rightarrow+\infty)$ it acts as a wall, marked by the vertical line, that separates the regions with negative and positive values of $\phi_c$.
		The blue point indicates the values of the scalar field and the potential at the present time.
	}
\end{figure}

In Fig.~\ref{plot1} we present the shape of the full potential obtained, as well as the approximation for large $a/a_0$ (see Eq.~\eqref{4.6}). We mark the present time values of the potential and scalar field by a blue point on the curve of the potential. From the results of the Planck mission \cite{Planck2013} we obtained the values $\Omega_{V_\infty}=0.6935$, $\Omega_\textrm{dust}=0.3065$, $\Omega_\textrm{rad}=9.117\times10^{-5}$. The value of $\Omega_\mathcal{K}+\Omega_\beta$ was set at a conservative value of $10^{-2}$. We would like to remind the reader that the observational constraints on the curvature term $\Omega_\mathcal{K}$ cannot be applied to a general term whose contribution to the energy density evolves as $1/a^2$. Therefore, $\Omega_\beta$ remains a free parameter constrained only by the requirement that $\Omega_\beta\ll \Omega_\textrm{dust},\Omega_{V_\infty}$.


\section{\label{sec:5}Conclusion}

In our paper, we have considered the late stage of the Universe evolution in the case when our Universe is filled with a minimal scalar field. We have also included dust-like matter (in the form of discrete distributed galaxies and groups of galaxies) and radiation as matter sources. To study the motion of galaxies in this Universe, we should know the distribution of the gravitational potential, which can be found from the theory of scalar perturbations. We considered this theory in the mechanical approach \cite{EZcosm2,EZcosm1}. In this approach, all types of inhomogeneities, e.g. galaxies as well as inhomogeneities associated with the fluctuations of other form of matter, have non-relativistic velocities. In this case, different types of inhomogeneities do not run away considerably during the Universe evolution. Moreover, fluctuations of the energy density of such perfect fluids are usually concentrated around the inhomogeneities of dust-like matter (i.e. galaxies). From this point, we call those perfect fluids "coupled" \cite{coupled}. They can screen the gravitational potential of galaxies \cite{BUZ1} and can also play the role of dark matter flattening the rotation curves of dwarf galaxies \cite{Laslo2}.

In the present paper, we have investigated the possibility for a scalar field to be coupled with galaxies in the late Universe. For such scalar fields to exist, we have shown that they have to meet certain conditions.
First, at the background level, such scalar field behaves as a two component perfect fluid: a network of frustrated cosmic strings with the EoS parameter $w=-1/3$ and a cosmological constant.
Second, it must have a certain form of the potential, which was determined and presented in Fig.~\ref{plot1}. Using the recent Planck data \cite{Planck2013} to constrain the parameters in our model, we find that the potential of the scalar field should be very flat at the present time. This flatness agrees with the asymptotic cosmological constant-like behaviour at the background level and leads the late time cosmic acceleration.
Third, the fluctuations of this field are absent but the fluctuations of the energy density and pressure are non-zero.
We have demonstrated that these fluctuations are concentrated around galaxies, in full agreement with the coupling condition. To show it, we have found the solutions for the gravitational potential. These solutions were obtained for all three possible spacial topologies of the Universe: flat, open and closed. All these gravitational potentials satisfy the important condition that their values averaged over the whole Universe are equal to zero. This leads to another physical reasonable result that fluctuations of the energy density of scalar field averaged over the Universe are also equal to zero. The main result of the paper is that the coupled scalar fields may exist under the conditions mentioned above and can provide the late cosmic acceleration.
It is worth to notice that inhomogeneous cosmological models can explain the coupling of inhomogeneities to a scalar field in a natural and effective way; i.e. without introducing a fundamental scalar field, please see Ref.~\cite{Buchert:2006ya}.

\section*{Acknowledgements}

A.Zh. acknowledges the hospitality of UBI during the completion of a part of this work.
The authors would like to thank Maxim Eingorn for his contribution during the initial phase of work and for the useful discussion of the obtained results.
J.M. is thankful to UPV/EHU for a PhD fellowship and UBI for the hospitality during the completion of part of this work and acknowledges the support from the Basque government Grant No. IT592-13 (Spain) and Fondos FEDER, under grant FIS2014-57956-P (Spanish Government).
The work of MBL is supported by the Portuguese Agency ``Funda\c{c}\~ao para a Ci\^encia e Tecnologia'' through an Investigador FCT Research contract, with reference IF/01442/2013/CP1196/CT0001. She also wishes to acknowledge the support from the Basque government Grant No. IT592-13 (Spain) and Fondos FEDER, under grant FIS2014-57956-P (Spanish Government).
K.S.K. is grateful for the support of the Grant SFRH/BD/51980/2012 from the Portuguese
Agency ``Funda\c{c}\~ao para a Ci\^encia e Tecnologia''.
This research work was supported by the Portuguese grant UID/MAT/00212/2013.



\end{document}